\documentclass[
reprint,
superscriptaddress,
amsmath,
amssymb,
aps,
pra,
longbibliography,
rsi,
raggedbottom]{revtex4-1}
\usepackage{color}
\usepackage{graphicx}
\usepackage[labelfont=bf]{caption}
\usepackage{subcaption}
\usepackage{dcolumn}
\usepackage{bm}
\usepackage{braket}
\usepackage{float}
\usepackage{multirow}
\usepackage{hhline}
\usepackage{mathtools}
\usepackage{qcircuit}
\usepackage{svg}
\usepackage{lipsum}
\usepackage{hyperref}
\hypersetup{colorlinks=true,linkcolor=blue,urlcolor=blue}

\begin{document}
\title{Diamond Circuits for Surface Codes}

\newcommand{\googla}{\affiliation{Google Quantum AI, Venice, CA 90291, USA}}

\author{Dripto M. Debroy}
\email{dripto@google.com}
\googla

\begin{abstract}
We present and benchmark an interesting circuit family which we call diamond circuits, that use a mid-cycle construction built around the subsystem surface code to implement a surface code on a Lieb or ``Heavy-Square'' lattice. This makes them more qubit- and measurement-efficient than previous constructions. These circuits are described via the LUCI framework, and are effectively circuits with half the measure qubits dropped out of the grid. These circuits preserve the spacelike distance of the code, but suffer a penalty in timelike distance, and could be useful in regimes where quantum computers are limited by frequency collisions or number of control lines. 
\end{abstract}

\maketitle

As quantum computers scale, different parameters may limit the sizes of systems, including fabrication capability, decoder throughput, physical error rates, and number of control lines. 
Qubit grids with lower connectivity have been used to mitigate the effects of frequency collision issues and crosstalk~\cite{gambetta2017building, chamberland2020topological, Benito2025comparativestudyof}. Specifically, the grid necessary for this circuit is the ``Heavy-Square'' or Lieb lattice, where two-thirds of the qubits only see two neighbors. Despite these promising experimental advantages, it was thought to not have a competitive quantum error correcting code, unlike the usual square lattice or the Heavy-Hex lattice. In this manuscript we show that the ``Heavy-Square'' lattice can implement a circuit whose end-cycle state is a surface code, one of the most well studied quantum error correcting codes~\cite{bravyi1998quantumcodeslatticeboundary, dennis2002topological, fowler2012surface, bravyi2013subsystem, acharya2024quantum}. 

\begin{figure}
    \centering
    \includegraphics[width=0.9\linewidth]{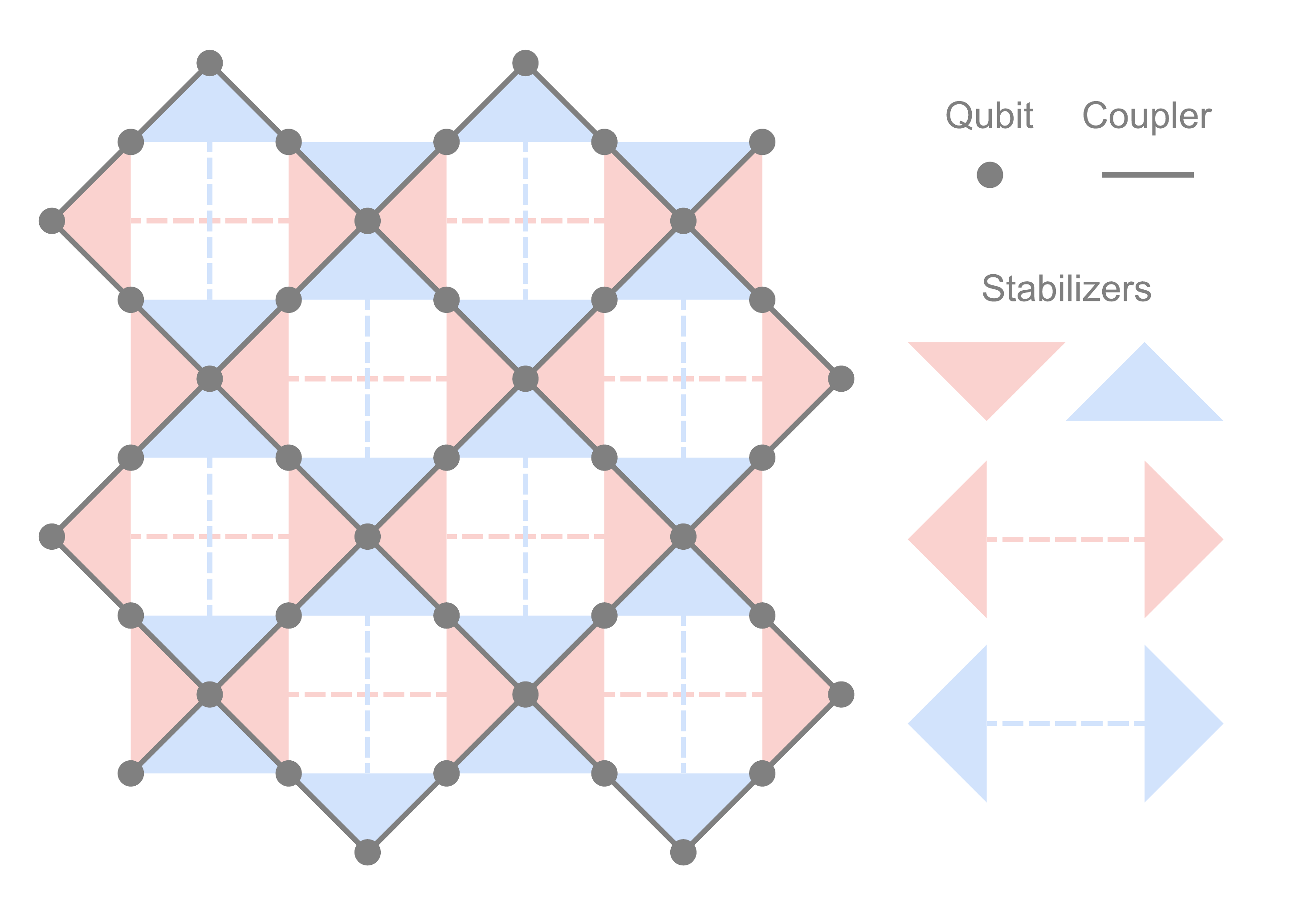}
    \caption{\textbf{Qubit grid and mid-cycle state for the distance-5 diamond circuit. } Qubit grid for a distance-5 diamond circuit with the mid-cycle gauge operators and stabilizers colored in pink and blue for $X$- and $Z$-type operators, respectively. Dotted lines indicated gauge operators which are paired to produce weight-6 stabilizers, while the boundaries have weight-3 stabilizers which satisfy commutation requirements without pairing.}
    \label{fig:grid}
\end{figure}

The circuit family we present is an instance of the LUCI framework~\cite{debroy2024lucisurfacecodedropouts} where the mid-cycle state is a subsystem surface code~\cite{bravyi2013subsystem}. The LUCI construction allows us to intentionally remove qubits from a surface code implementation, letting us construct a distance-$d$ surface code using roughly $1.5d^2$ qubits, compared to the $2d^2 - 1$ qubits of the standard rotated surface code circuit. An architecture built on these circuits could be easier to fabricate, and would require fewer control lines than the one designed around the usual circuit, allowing for a larger distance in a line-limited context. Fig.~\ref{fig:grid} shows this lattice, where qubits exist on vertices and couplers are shown as gray lines connecting neighboring qubits. The colored triangles show the stabilizer and gauge operators of the mid-cycle state. When the gauges are paired according to the dotted lines of the same color, they form superstabilizers. We refer to these circuits as diamond circuits due to the resemblance of this mid-cycle state to the facets of a cut diamond. The approach to building error correction circuits around the mid-cycle state was introduced in Ref.~\cite{mcewen2023relaxing} and further explored in Refs.~\cite{gidney2023new, shaw2024loweringconnectivityrequirementsbivariate}, while the LUCI framework which we will use to describe this particular family of circuits is fully outlined in Ref.~\cite{debroy2024lucisurfacecodedropouts}. We will briefly summarize these ideas to introduce the circuits, before focusing on numerical benchmarking and comparisons to the standard surface code implementation.

The idea of building circuits around the mid-cycle state of a code begins with the observation that halfway between measurements of the usual surface code circuit, the stabilizers of the code propagate into an unrotated surface code supported on the data and measure qubits. The key insight in Ref.~\cite{mcewen2023relaxing} is that one can build valid fault-tolerant circuits by measuring these mid-cycle stabilizers in place, before then returning to the mid-cycle state. This is further developed in Ref.~\cite{debroy2024lucisurfacecodedropouts}, where the possible space of mid-cycle states is expanded for the purpose of adapting surface codes to broken components. The general idea remains, with mid-cycle stabilizers and gauge operators being measured in place and then reset before being returned to their original footprint.

\begin{figure*}
    \centering
    \includegraphics[width=0.95\linewidth]{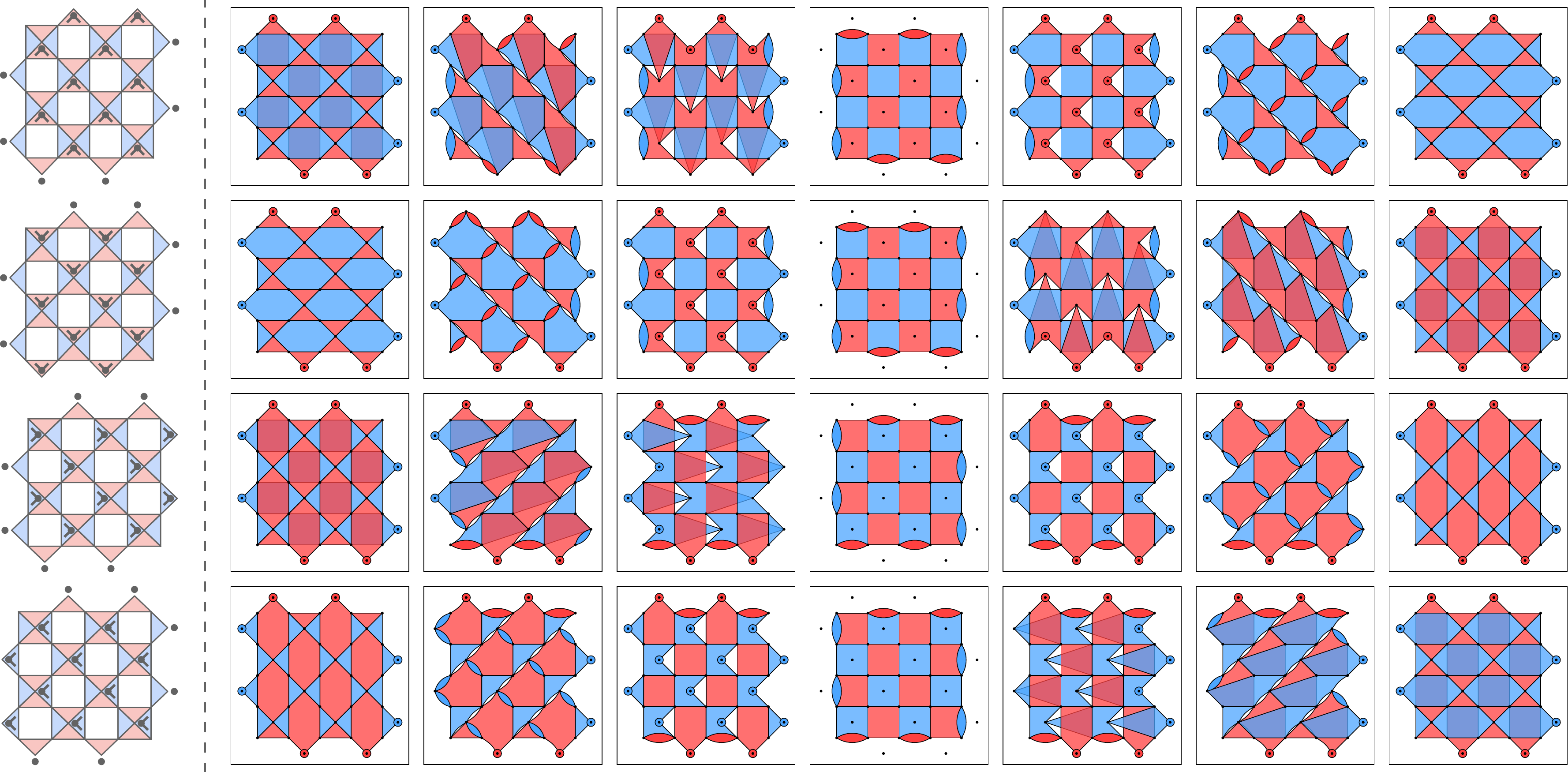}
    \caption{\textbf{LUCI Diagram and detector slices for diamond circuits. } The left side of the figure shows a LUCI diagram for measuring a distance-5 diamond circuit, while the panels to the right of the dotted line show timelike slices of the detectors after every layer of operations. The first two layers are sets of CNOT operations, after which information is measured and qubits are reset before the state is returned to the mid-cycle. From the overlaps of the weight-6 operations we can follow how the first round has one of the weight-3 X-type gauge operators, shown in red, measured for each stabilizer. After the reset we now have the weight-6 stabilizer split into its component weight-3 gauges, indicating that there is one detecting region closing while the next one has already started opening. Since the rounds go from mid-cycle to mid-cycle, the first and last panels of each detector slice show the mid-cycle state from Fig.~\ref{fig:grid}, while the middle panels show traditional surface code end-cycle state, where the stabilizers shift around such that the X-type operators can be measured in the first two rounds and the Z-type operators in the last two rounds. \href{https://algassert.com/crumble\#circuit=Q(0,2)0;Q(0,6)1;Q(1,1)2;Q(1,3)3;Q(1,5)4;Q(1,7)5;Q(1,9)6;Q(2,0)7;Q(2,4)8;Q(2,8)9;Q(3,1)10;Q(3,3)11;Q(3,5)12;Q(3,7)13;Q(3,9)14;Q(4,2)15;Q(4,6)16;Q(4,10)17;Q(5,1)18;Q(5,3)19;Q(5,5)20;Q(5,7)21;Q(5,9)22;Q(6,0)23;Q(6,4)24;Q(6,8)25;Q(7,1)26;Q(7,3)27;Q(7,5)28;Q(7,7)29;Q(7,9)30;Q(8,2)31;Q(8,6)32;Q(8,10)33;Q(9,1)34;Q(9,3)35;Q(9,5)36;Q(9,7)37;Q(9,9)38;Q(10,4)39;Q(10,8)40;POLYGON(0,0,1,0.25)2_3_0;POLYGON(0,0,1,0.25)10_15_11;POLYGON(0,0,1,0.25)18_19_15;POLYGON(0,0,1,0.25)26_31_27;POLYGON(0,0,1,0.25)34_35_31;POLYGON(0,0,1,0.25)3_8_4;POLYGON(0,0,1,0.25)11_12_8;POLYGON(0,0,1,0.25)12_16_13;POLYGON(0,0,1,0.25)20_21_16;POLYGON(0,0,1,0.25)21_25_22;POLYGON(0,0,1,0.25)29_30_25;POLYGON(0,0,1,0.25)37_40_38;POLYGON(0,0,1,0.25)36_37_32;POLYGON(0,0,1,0.25)35_39_36;POLYGON(0,0,1,0.25)27_28_24;POLYGON(0,0,1,0.25)28_32_29;POLYGON(0,0,1,0.25)19_24_20;POLYGON(0,0,1,0.25)4_5_1;POLYGON(0,0,1,0.25)13_14_9;POLYGON(0,0,1,0.25)5_9_6;POLYGON(1,0,0,0.25)7_10_2;POLYGON(1,0,0,0.25)3_11_8;POLYGON(1,0,0,0.25)8_12_4;POLYGON(1,0,0,0.25)5_13_9;POLYGON(1,0,0,0.25)12_20_16;POLYGON(1,0,0,0.25)15_19_11;POLYGON(1,0,0,0.25)23_26_18;POLYGON(1,0,0,0.25)10_18_15;POLYGON(1,0,0,0.25)19_27_24;POLYGON(1,0,0,0.25)24_28_20;POLYGON(1,0,0,0.25)21_29_25;POLYGON(1,0,0,0.25)16_21_13;POLYGON(1,0,0,0.25)14_22_17;POLYGON(1,0,0,0.25)30_38_33;POLYGON(1,0,0,0.25)32_37_29;POLYGON(1,0,0,0.25)28_36_32;POLYGON(1,0,0,0.25)31_35_27;POLYGON(1,0,0,0.25)26_34_31;POLYGON(1,0,0,0.25)9_14_6;POLYGON(1,0,0,0.25)25_30_22;TICK;R_0_40_39_1;RX_2_3_10_4_11_18_5_12_19_26_6_13_20_27_34_14_21_28_35_22_29_36_30_37_38_32_7_33_8_15_23_9_16_24_31_17_25;TICK;CX_24_20_25_22_7_2_23_18_9_6_32_29_31_27_8_4_15_11_16_13;TICK;CX_7_10_16_21_25_30_8_12_31_35_9_14_32_37_15_19_23_26_24_28;TICK;CX_38_40_29_32_36_39_13_16_27_31_4_8_11_15_22_25_6_9_20_24;TICK;CX_5_9_10_15_3_8_26_31_28_32_35_39_12_16_37_40_19_24_21_25;TICK;M_0_32_39_1_8_40_15_9_16_24_31_25;MX_7_23_17_33;DT(0,2,0)rec[-16];DT(0,6,0)rec[-13];DT(2,0,0)rec[-4];DT(6,0,0)rec[-3];DT(4,10,0)rec[-2];DT(8,10,0)rec[-1];TICK;R_0_32_39_1_8_40_15_9_16_24_31_25;RX_7_23_17_33;TICK;CX_5_9_10_15_3_8_26_31_28_32_35_39_12_16_37_40_19_24_21_25;TICK;CX_38_40_29_32_36_39_13_16_27_31_4_8_11_15_22_25_6_9_20_24;TICK;TICK;CX_29_25_36_32_13_9_27_24_18_15_4_1_34_31_11_8_20_16_2_0;TICK;CX_35_31_28_24_14_9_12_8_3_0_19_15_21_16_5_1_37_32_30_25;TICK;M_0_32_39_1_8_40_15_9_16_24_31_25;MX_7_23_17_33;DT(10,4,1)rec[-14]_rec[-30];DT(10,8,1)rec[-11]_rec[-27];DT(2,0,1)rec[-4];DT(6,0,1)rec[-3];DT(4,10,1)rec[-2];DT(8,10,1)rec[-1];TICK;R_0_32_39_1_8_40_15_9_16_24_31_25;RX_7_23_17_33;TICK;CX_35_31_28_24_14_9_12_8_3_0_19_15_21_16_5_1_37_32_30_25;TICK;CX_29_25_36_32_13_9_27_24_18_15_4_1_34_31_11_8_20_16_2_0;TICK;TICK;CX_33_30_25_21_8_3_17_14_16_12_9_5_31_26_24_19_32_28_15_10;TICK;CX_15_18_33_38_16_20_24_27_25_29_32_36_9_13_31_34_8_11_17_22;TICK;M_0_40_39_1;MX_32_7_33_8_15_23_9_16_24_31_17_25;DT(0,2,2)rec[-16]_rec[-32];DT(10,8,2)rec[-15];DT(10,4,2)rec[-14];DT(0,6,2)rec[-13]_rec[-29];DT(2,0,2)rec[-11];DT(4,2,2)rec[-8];DT(6,0,2)rec[-7];DT(8,2,2)rec[-3];TICK;R_0_40_39_1;RX_32_7_33_8_15_23_9_16_24_31_17_25;TICK;CX_15_18_33_38_16_20_24_27_25_29_32_36_9_13_31_34_8_11_17_22;TICK;CX_33_30_25_21_8_3_17_14_16_12_9_5_31_26_24_19_32_28_15_10;TICK;TICK;CX_7_10_16_21_25_30_8_12_31_35_9_14_32_37_15_19_23_26_24_28;TICK;CX_24_20_25_22_7_2_23_18_9_6_32_29_31_27_8_4_15_11_16_13;TICK;M_0_40_39_1;MX_32_7_33_8_15_23_9_16_24_31_17_25;DT(0,2,3)rec[-16];DT(10,8,3)rec[-15];DT(10,4,3)rec[-14];DT(0,6,3)rec[-13];DT(8,10,3)rec[-12]_rec[-26]_rec[-28]_rec[-33]_rec[-49];DT(2,0,3)rec[-11]_rec[-25]_rec[-27]_rec[-36]_rec[-52];DT(8,10,4)rec[-10]_rec[-26];DT(2,8,3)rec[-9]_rec[-22]_rec[-25];DT(4,6,3)rec[-8]_rec[-21]_rec[-24];DT(6,0,3)rec[-7]_rec[-20]_rec[-23]_rec[-35]_rec[-51];DT(2,8,4)rec[-6]_rec[-22];DT(4,10,3)rec[-5]_rec[-18]_rec[-21]_rec[-34]_rec[-50];DT(6,8,3)rec[-4]_rec[-17]_rec[-20];DT(8,2,3)rec[-3]_rec[-19]_rec[-28];DT(4,10,4)rec[-2]_rec[-18];DT(6,8,4)rec[-1]_rec[-17];TICK;R_0_40_39_1;RX_32_7_33_8_15_23_9_16_24_31_17_25;TICK;CX_24_20_25_22_7_2_23_18_9_6_32_29_31_27_8_4_15_11_16_13;TICK;CX_7_10_16_21_25_30_8_12_31_35_9_14_32_37_15_19_23_26_24_28;TICK;TICK;CX_38_40_29_32_36_39_13_16_27_31_4_8_11_15_22_25_6_9_20_24;TICK;CX_5_9_10_15_3_8_26_31_28_32_35_39_12_16_37_40_19_24_21_25;TICK;M_0_32_39_1_8_40_15_9_16_24_31_25;MX_7_23_17_33;DT(0,2,5)rec[-16];DT(0,6,5)rec[-13];DT(2,4,5)rec[-12]_rec[-60];DT(2,8,5)rec[-9]_rec[-57];DT(2,0,5)rec[-4]_rec[-27];DT(6,0,5)rec[-3]_rec[-23];DT(4,10,5)rec[-2];DT(8,10,5)rec[-1];TICK;R_0_32_39_1_8_40_15_9_16_24_31_25;RX_7_23_17_33;TICK;CX_5_9_10_15_3_8_26_31_28_32_35_39_12_16_37_40_19_24_21_25;TICK;CX_38_40_29_32_36_39_13_16_27_31_4_8_11_15_22_25_6_9_20_24;TICK;TICK;CX_29_25_36_32_13_9_27_24_18_15_4_1_34_31_11_8_20_16_2_0;TICK;CX_35_31_28_24_14_9_12_8_3_0_19_15_21_16_5_1_37_32_30_25;TICK;M_0_32_39_1_8_40_15_9_16_24_31_25;MX_7_23_17_33;DT(4,2,6)rec[-16]_rec[-26]_rec[-32]_rec[-48]_rec[-64]_rec[-74];DT(8,6,6)rec[-15]_rec[-31];DT(10,4,6)rec[-14]_rec[-30];DT(4,6,6)rec[-13]_rec[-24]_rec[-29]_rec[-45]_rec[-61]_rec[-72];DT(6,4,6)rec[-12]_rec[-23]_rec[-28]_rec[-71];DT(10,8,6)rec[-11]_rec[-27];DT(8,2,6)rec[-10]_rec[-22]_rec[-26]_rec[-70];DT(6,8,6)rec[-9]_rec[-21]_rec[-25]_rec[-69];DT(6,6,6)rec[-8]_rec[-24]_rec[-31]_rec[-79];DT(8.5,4,6)rec[-7]_rec[-23]_rec[-30]_rec[-46]_rec[-62]_rec[-78];DT(8,2,7)rec[-6]_rec[-22];DT(8.5,8,6)rec[-5]_rec[-21]_rec[-27]_rec[-47]_rec[-63]_rec[-75];DT(2,0,6)rec[-4];DT(6,0,6)rec[-3];DT(4,10,6)rec[-2];DT(8,10,6)rec[-1];TICK;R_0_32_39_1_8_40_15_9_16_24_31_25;RX_7_23_17_33;TICK;CX_35_31_28_24_14_9_12_8_3_0_19_15_21_16_5_1_37_32_30_25;TICK;CX_29_25_36_32_13_9_27_24_18_15_4_1_34_31_11_8_20_16_2_0;TICK;TICK;CX_33_30_25_21_8_3_17_14_16_12_9_5_31_26_24_19_32_28_15_10;TICK;CX_15_18_33_38_16_20_24_27_25_29_32_36_9_13_31_34_8_11_17_22;TICK;M_0_40_39_1;MX_32_7_33_8_15_23_9_16_24_31_17_25;DT(0,2,8)rec[-16]_rec[-32];DT(10,8,8)rec[-15];DT(10,4,8)rec[-14];DT(0,6,8)rec[-13]_rec[-29];DT(2,0,8)rec[-11];DT(4,2,8)rec[-8]_rec[-56];DT(6,0,8)rec[-7];DT(8,2,8)rec[-3]_rec[-51];TICK;R_0_40_39_1;RX_32_7_33_8_15_23_9_16_24_31_17_25;TICK;CX_15_18_33_38_16_20_24_27_25_29_32_36_9_13_31_34_8_11_17_22;TICK;CX_33_30_25_21_8_3_17_14_16_12_9_5_31_26_24_19_32_28_15_10;TICK;TICK;CX_7_10_16_21_25_30_8_12_31_35_9_14_32_37_15_19_23_26_24_28;TICK;CX_24_20_25_22_7_2_23_18_9_6_32_29_31_27_8_4_15_11_16_13;TICK;M_0_40_39_1;MX_32_7_33_8_15_23_9_16_24_31_17_25;DT(0,2,9)rec[-16];DT(10,8,9)rec[-15];DT(10,4,9)rec[-14];DT(0,6,9)rec[-13];DT(8,10,9)rec[-12]_rec[-26]_rec[-28]_rec[-33]_rec[-49]_rec[-74];DT(2,4,9)rec[-11]_rec[-25]_rec[-27]_rec[-36]_rec[-52]_rec[-73];DT(8,10,10)rec[-10]_rec[-26];DT(2,8,9)rec[-9]_rec[-22]_rec[-25]_rec[-70];DT(4,6,9)rec[-8]_rec[-21]_rec[-24]_rec[-69];DT(6,4,9)rec[-7]_rec[-20]_rec[-23]_rec[-35]_rec[-51]_rec[-68];DT(2,8,10)rec[-6]_rec[-22];DT(4,10,9)rec[-5]_rec[-18]_rec[-21]_rec[-34]_rec[-50]_rec[-66];DT(6,8,9)rec[-4]_rec[-17]_rec[-20]_rec[-65];DT(8,6,9)rec[-3]_rec[-19]_rec[-28]_rec[-76];DT(4,10,10)rec[-2]_rec[-18];DT(6,8,10)rec[-1]_rec[-17];TICK;R_0_40_39_1;RX_32_7_33_8_15_23_9_16_24_31_17_25;TICK;CX_24_20_25_22_7_2_23_18_9_6_32_29_31_27_8_4_15_11_16_13;TICK;CX_7_10_16_21_25_30_8_12_31_35_9_14_32_37_15_19_23_26_24_28;TICK;TICK;CX_38_40_29_32_36_39_13_16_27_31_4_8_11_15_22_25_6_9_20_24;TICK;CX_5_9_10_15_3_8_26_31_28_32_35_39_12_16_37_40_19_24_21_25;TICK;M_0_32_39_1_8_40_15_9_16_24_31_25;MX_7_23_17_33;DT(0,2,11)rec[-16];DT(0,6,11)rec[-13];DT(2,4,11)rec[-12]_rec[-60];DT(2,8,11)rec[-9]_rec[-57];DT(2,0,11)rec[-4]_rec[-27];DT(6,0,11)rec[-3]_rec[-23];DT(4,10,11)rec[-2];DT(8,10,11)rec[-1];TICK;R_0_32_39_1_8_40_15_9_16_24_31_25;RX_7_23_17_33;TICK;CX_5_9_10_15_3_8_26_31_28_32_35_39_12_16_37_40_19_24_21_25;TICK;CX_38_40_29_32_36_39_13_16_27_31_4_8_11_15_22_25_6_9_20_24;TICK;TICK;CX_29_25_36_32_13_9_27_24_18_15_4_1_34_31_11_8_20_16_2_0;TICK;CX_35_31_28_24_14_9_12_8_3_0_19_15_21_16_5_1_37_32_30_25;TICK;M_0_32_39_1_8_40_15_9_16_24_31_25;MX_7_23_17_33;DT(4,2,12)rec[-16]_rec[-26]_rec[-32]_rec[-48]_rec[-64]_rec[-74];DT(8,6,12)rec[-15]_rec[-31];DT(10,4,12)rec[-14]_rec[-30];DT(4,6,12)rec[-13]_rec[-24]_rec[-29]_rec[-45]_rec[-61]_rec[-72];DT(6,4,12)rec[-12]_rec[-23]_rec[-28]_rec[-71];DT(10,8,12)rec[-11]_rec[-27];DT(8,2,12)rec[-10]_rec[-22]_rec[-26]_rec[-70];DT(6,8,12)rec[-9]_rec[-21]_rec[-25]_rec[-69];DT(6,6,12)rec[-8]_rec[-24]_rec[-31]_rec[-79];DT(8.5,4,12)rec[-7]_rec[-23]_rec[-30]_rec[-46]_rec[-62]_rec[-78];DT(8,2,13)rec[-6]_rec[-22];DT(8.5,8,12)rec[-5]_rec[-21]_rec[-27]_rec[-47]_rec[-63]_rec[-75];DT(2,0,12)rec[-4];DT(6,0,12)rec[-3];DT(4,10,12)rec[-2];DT(8,10,12)rec[-1];TICK;R_0_32_39_1_8_40_15_9_16_24_31_25;RX_7_23_17_33;TICK;CX_35_31_28_24_14_9_12_8_3_0_19_15_21_16_5_1_37_32_30_25;TICK;CX_29_25_36_32_13_9_27_24_18_15_4_1_34_31_11_8_20_16_2_0;TICK;TICK;CX_33_30_25_21_8_3_17_14_16_12_9_5_31_26_24_19_32_28_15_10;TICK;CX_15_18_33_38_16_20_24_27_25_29_32_36_9_13_31_34_8_11_17_22;TICK;M_0_40_39_1;MX_32_7_33_8_15_23_9_16_24_31_17_25;DT(0,2,14)rec[-16]_rec[-32];DT(10,8,14)rec[-15];DT(10,4,14)rec[-14];DT(0,6,14)rec[-13]_rec[-29];DT(2,0,14)rec[-11];DT(4,2,14)rec[-8]_rec[-56];DT(6,0,14)rec[-7];DT(8,2,14)rec[-3]_rec[-51];TICK;R_0_40_39_1;RX_32_7_33_8_15_23_9_16_24_31_17_25;TICK;CX_15_18_33_38_16_20_24_27_25_29_32_36_9_13_31_34_8_11_17_22;TICK;CX_33_30_25_21_8_3_17_14_16_12_9_5_31_26_24_19_32_28_15_10;TICK;TICK;CX_7_10_16_21_25_30_8_12_31_35_9_14_32_37_15_19_23_26_24_28;TICK;CX_24_20_25_22_7_2_23_18_9_6_32_29_31_27_8_4_15_11_16_13;TICK;M_0_40_39_1;MX_32_7_33_8_15_23_9_16_24_31_17_25;DT(0,2,15)rec[-16];DT(10,8,15)rec[-15];DT(10,4,15)rec[-14];DT(0,6,15)rec[-13];DT(8,10,15)rec[-12]_rec[-26]_rec[-28]_rec[-33]_rec[-49]_rec[-74];DT(2,4,15)rec[-11]_rec[-25]_rec[-27]_rec[-36]_rec[-52]_rec[-73];DT(8,10,16)rec[-10]_rec[-26];DT(2,8,15)rec[-9]_rec[-22]_rec[-25]_rec[-70];DT(4,6,15)rec[-8]_rec[-21]_rec[-24]_rec[-69];DT(6,4,15)rec[-7]_rec[-20]_rec[-23]_rec[-35]_rec[-51]_rec[-68];DT(2,8,16)rec[-6]_rec[-22];DT(4,10,15)rec[-5]_rec[-18]_rec[-21]_rec[-34]_rec[-50]_rec[-66];DT(6,8,15)rec[-4]_rec[-17]_rec[-20]_rec[-65];DT(8,6,15)rec[-3]_rec[-19]_rec[-28]_rec[-76];DT(4,10,16)rec[-2]_rec[-18];DT(6,8,16)rec[-1]_rec[-17];TICK;R_0_40_39_1;RX_32_7_33_8_15_23_9_16_24_31_17_25;TICK;CX_24_20_25_22_7_2_23_18_9_6_32_29_31_27_8_4_15_11_16_13;TICK;CX_7_10_16_21_25_30_8_12_31_35_9_14_32_37_15_19_23_26_24_28;TICK;TICK;CX_38_40_29_32_36_39_13_16_27_31_4_8_11_15_22_25_6_9_20_24;TICK;CX_5_9_10_15_3_8_26_31_28_32_35_39_12_16_37_40_19_24_21_25;TICK;M_0_32_39_1_8_40_15_9_16_24_31_25;MX_7_23_17_33;DT(0,2,17)rec[-16];DT(0,6,17)rec[-13];DT(2,4,17)rec[-12]_rec[-60];DT(2,8,17)rec[-9]_rec[-57];DT(2,0,17)rec[-4]_rec[-27];DT(6,0,17)rec[-3]_rec[-23];DT(4,10,17)rec[-2];DT(8,10,17)rec[-1];TICK;R_0_32_39_1_8_40_15_9_16_24_31_25;RX_7_23_17_33;TICK;CX_5_9_10_15_3_8_26_31_28_32_35_39_12_16_37_40_19_24_21_25;TICK;CX_38_40_29_32_36_39_13_16_27_31_4_8_11_15_22_25_6_9_20_24;TICK;TICK;CX_29_25_36_32_13_9_27_24_18_15_4_1_34_31_11_8_20_16_2_0;TICK;CX_35_31_28_24_14_9_12_8_3_0_19_15_21_16_5_1_37_32_30_25;TICK;M_0_32_39_1_8_40_15_9_16_24_31_25;MX_7_23_17_33;DT(4,2,18)rec[-16]_rec[-26]_rec[-32]_rec[-48]_rec[-64]_rec[-74];DT(8,6,18)rec[-15]_rec[-31];DT(10,4,18)rec[-14]_rec[-30];DT(4,6,18)rec[-13]_rec[-24]_rec[-29]_rec[-45]_rec[-61]_rec[-72];DT(6,4,18)rec[-12]_rec[-23]_rec[-28]_rec[-71];DT(10,8,18)rec[-11]_rec[-27];DT(8,2,18)rec[-10]_rec[-22]_rec[-26]_rec[-70];DT(6,8,18)rec[-9]_rec[-21]_rec[-25]_rec[-69];DT(6,6,18)rec[-8]_rec[-24]_rec[-31]_rec[-79];DT(8.5,4,18)rec[-7]_rec[-23]_rec[-30]_rec[-46]_rec[-62]_rec[-78];DT(8,2,19)rec[-6]_rec[-22];DT(8.5,8,18)rec[-5]_rec[-21]_rec[-27]_rec[-47]_rec[-63]_rec[-75];DT(2,0,18)rec[-4];DT(6,0,18)rec[-3];DT(4,10,18)rec[-2];DT(8,10,18)rec[-1];TICK;R_0_32_39_1_8_40_15_9_16_24_31_25;RX_7_23_17_33;TICK;CX_35_31_28_24_14_9_12_8_3_0_19_15_21_16_5_1_37_32_30_25;TICK;CX_29_25_36_32_13_9_27_24_18_15_4_1_34_31_11_8_20_16_2_0;TICK;TICK;CX_33_30_25_21_8_3_17_14_16_12_9_5_31_26_24_19_32_28_15_10;TICK;CX_15_18_33_38_16_20_24_27_25_29_32_36_9_13_31_34_8_11_17_22;TICK;M_0_40_39_1;MX_32_7_33_8_15_23_9_16_24_31_17_25;DT(0,2,20)rec[-16]_rec[-32];DT(10,8,20)rec[-15];DT(10,4,20)rec[-14];DT(0,6,20)rec[-13]_rec[-29];DT(2,0,20)rec[-11];DT(4,2,20)rec[-8]_rec[-56];DT(6,0,20)rec[-7];DT(8,2,20)rec[-3]_rec[-51];TICK;R_0_40_39_1;RX_32_7_33_8_15_23_9_16_24_31_17_25;TICK;CX_15_18_33_38_16_20_24_27_25_29_32_36_9_13_31_34_8_11_17_22;TICK;CX_33_30_25_21_8_3_17_14_16_12_9_5_31_26_24_19_32_28_15_10;TICK;TICK;CX_7_10_16_21_25_30_8_12_31_35_9_14_32_37_15_19_23_26_24_28;TICK;CX_24_20_25_22_7_2_23_18_9_6_32_29_31_27_8_4_15_11_16_13;TICK;M_0_40_39_1;MX_32_7_33_8_15_23_9_16_24_31_17_25;DT(0,2,21)rec[-16];DT(10,8,21)rec[-15];DT(10,4,21)rec[-14];DT(0,6,21)rec[-13];DT(8,10,21)rec[-12]_rec[-26]_rec[-28]_rec[-33]_rec[-49]_rec[-74];DT(2,4,21)rec[-11]_rec[-25]_rec[-27]_rec[-36]_rec[-52]_rec[-73];DT(8,10,22)rec[-10]_rec[-26];DT(2,8,21)rec[-9]_rec[-22]_rec[-25]_rec[-70];DT(4,6,21)rec[-8]_rec[-21]_rec[-24]_rec[-69];DT(6,4,21)rec[-7]_rec[-20]_rec[-23]_rec[-35]_rec[-51]_rec[-68];DT(2,8,22)rec[-6]_rec[-22];DT(4,10,21)rec[-5]_rec[-18]_rec[-21]_rec[-34]_rec[-50]_rec[-66];DT(6,8,21)rec[-4]_rec[-17]_rec[-20]_rec[-65];DT(8,6,21)rec[-3]_rec[-19]_rec[-28]_rec[-76];DT(4,10,22)rec[-2]_rec[-18];DT(6,8,22)rec[-1]_rec[-17];TICK;R_0_40_39_1;RX_32_7_33_8_15_23_9_16_24_31_17_25;TICK;CX_24_20_25_22_7_2_23_18_9_6_32_29_31_27_8_4_15_11_16_13;TICK;CX_7_10_16_21_25_30_8_12_31_35_9_14_32_37_15_19_23_26_24_28;TICK;TICK;CX_38_40_29_32_36_39_13_16_27_31_4_8_11_15_22_25_6_9_20_24;TICK;CX_5_9_10_15_3_8_26_31_28_32_35_39_12_16_37_40_19_24_21_25;TICK;M_0_32_39_1_8_40_15_9_16_24_31_25;MX_7_23_17_33;DT(0,2,23)rec[-16];DT(0,6,23)rec[-13];DT(2,4,23)rec[-12]_rec[-60];DT(2,8,23)rec[-9]_rec[-57];DT(2,0,23)rec[-4]_rec[-27];DT(6,0,23)rec[-3]_rec[-23];DT(4,10,23)rec[-2];DT(8,10,23)rec[-1];TICK;R_0_32_39_1_8_40_15_9_16_24_31_25;RX_7_23_17_33;TICK;CX_5_9_10_15_3_8_26_31_28_32_35_39_12_16_37_40_19_24_21_25;TICK;CX_38_40_29_32_36_39_13_16_27_31_4_8_11_15_22_25_6_9_20_24;TICK;TICK;CX_29_25_36_32_13_9_27_24_18_15_4_1_34_31_11_8_20_16_2_0;TICK;CX_35_31_28_24_14_9_12_8_3_0_19_15_21_16_5_1_37_32_30_25;TICK;M_0_32_39_1_8_40_15_9_16_24_31_25;MX_7_23_17_33;DT(4,2,24)rec[-16]_rec[-26]_rec[-32]_rec[-48]_rec[-64]_rec[-74];DT(8,6,24)rec[-15]_rec[-31];DT(10,4,24)rec[-14]_rec[-30];DT(4,6,24)rec[-13]_rec[-24]_rec[-29]_rec[-45]_rec[-61]_rec[-72];DT(6,4,24)rec[-12]_rec[-23]_rec[-28]_rec[-71];DT(10,8,24)rec[-11]_rec[-27];DT(8,2,24)rec[-10]_rec[-22]_rec[-26]_rec[-70];DT(6,8,24)rec[-9]_rec[-21]_rec[-25]_rec[-69];DT(6,6,24)rec[-8]_rec[-24]_rec[-31]_rec[-79];DT(8.5,4,24)rec[-7]_rec[-23]_rec[-30]_rec[-46]_rec[-62]_rec[-78];DT(8,2,25)rec[-6]_rec[-22];DT(8.5,8,24)rec[-5]_rec[-21]_rec[-27]_rec[-47]_rec[-63]_rec[-75];DT(2,0,24)rec[-4];DT(6,0,24)rec[-3];DT(4,10,24)rec[-2];DT(8,10,24)rec[-1];TICK;R_0_32_39_1_8_40_15_9_16_24_31_25;RX_7_23_17_33;TICK;CX_35_31_28_24_14_9_12_8_3_0_19_15_21_16_5_1_37_32_30_25;TICK;CX_29_25_36_32_13_9_27_24_18_15_4_1_34_31_11_8_20_16_2_0;TICK;TICK;CX_33_30_25_21_8_3_17_14_16_12_9_5_31_26_24_19_32_28_15_10;TICK;CX_15_18_33_38_16_20_24_27_25_29_32_36_9_13_31_34_8_11_17_22;TICK;M_0_40_39_1;MX_32_7_33_8_15_23_9_16_24_31_17_25;DT(0,2,26)rec[-16]_rec[-32];DT(10,8,26)rec[-15];DT(10,4,26)rec[-14];DT(0,6,26)rec[-13]_rec[-29];DT(2,0,26)rec[-11];DT(4,2,26)rec[-8]_rec[-56];DT(6,0,26)rec[-7];DT(8,2,26)rec[-3]_rec[-51];TICK;R_0_40_39_1;RX_32_7_33_8_15_23_9_16_24_31_17_25;TICK;CX_15_18_33_38_16_20_24_27_25_29_32_36_9_13_31_34_8_11_17_22;TICK;CX_33_30_25_21_8_3_17_14_16_12_9_5_31_26_24_19_32_28_15_10;TICK;TICK;CX_7_10_16_21_25_30_8_12_31_35_9_14_32_37_15_19_23_26_24_28;TICK;CX_24_20_25_22_7_2_23_18_9_6_32_29_31_27_8_4_15_11_16_13;TICK;M_0_40_39_1;MX_32_7_33_8_15_23_9_16_24_31_17_25_2_3_10_4_11_18_5_12_19_26_6_13_20_27_34_14_21_28_35_22_29_36_30_37_38;DT(0,2,27)rec[-41];DT(10,8,27)rec[-40];DT(10,4,27)rec[-39];DT(0,6,27)rec[-38];DT(8,10,27)rec[-37]_rec[-51]_rec[-53]_rec[-58]_rec[-74]_rec[-99];DT(2,4,27)rec[-36]_rec[-50]_rec[-52]_rec[-61]_rec[-77]_rec[-98];DT(8,10,28)rec[-35]_rec[-51];DT(2,8,27)rec[-34]_rec[-47]_rec[-50]_rec[-95];DT(4,6,27)rec[-33]_rec[-46]_rec[-49]_rec[-94];DT(6,4,27)rec[-32]_rec[-45]_rec[-48]_rec[-60]_rec[-76]_rec[-93];DT(2,8,28)rec[-31]_rec[-47];DT(4,10,27)rec[-30]_rec[-43]_rec[-46]_rec[-59]_rec[-75]_rec[-91];DT(6,8,27)rec[-29]_rec[-42]_rec[-45]_rec[-90];DT(8,6,27)rec[-28]_rec[-44]_rec[-53]_rec[-101];DT(4,10,28)rec[-27]_rec[-43];DT(6,8,28)rec[-26]_rec[-42];DT(3,1,27)rec[-23]_rec[-25]_rec[-36];DT(3,5,27)rec[-18]_rec[-21]_rec[-22]_rec[-24]_rec[-34];DT(5,3,27)rec[-17]_rec[-20]_rec[-21]_rec[-23]_rec[-33];DT(7,1,27)rec[-16]_rec[-20]_rec[-32];DT(3,9,27)rec[-10]_rec[-14]_rec[-15]_rec[-19]_rec[-31];DT(5,7,27)rec[-9]_rec[-13]_rec[-14]_rec[-18]_rec[-30];DT(7,5,27)rec[-8]_rec[-12]_rec[-13]_rec[-17]_rec[-29];DT(9,3,27)rec[-7]_rec[-11]_rec[-12]_rec[-16]_rec[-28];DT(5,9,27)rec[-6]_rec[-10]_rec[-27];DT(7,9,27)rec[-3]_rec[-5]_rec[-6]_rec[-9]_rec[-26];DT(9,7,27)rec[-2]_rec[-4]_rec[-5]_rec[-8]_rec[-37];DT(9,9,27)rec[-1]_rec[-3]_rec[-35];OI(0)rec[-15]_rec[-19]_rec[-22]_rec[-24]_rec[-25]}{Crumble link for a distance-5 circuit diamond circuit memory experiment.}}
    \label{fig:detslices}
\end{figure*}
In Fig.~\ref{fig:detslices} we show the LUCI diagram that describes a distance-5 diamond circuit, as well as time slices of the detecting regions of the circuit as the state propagates through the circuit. The rounds of the LUCI diagram, shown on the left of the plot, use gray shapes and dots to indicate which stabilizers and gauge operators are being measured and reprepared in that round. The gray lines in each shape indicate which two-qubit gates are used, while the dots indicate the qubits that are measured and reset. For more details on the exact compilations of these shapes into operators, see Figures 3 and 4 in Ref.~\cite{debroy2024lucisurfacecodedropouts}. The detector slices are laid out from mid-cycle to mid-cycle, making it easy to see that despite the mid-cycle state being a subsystem surface code~\cite{bravyi2013subsystem}, the end-cycle state is still the usual surface code state. In each round, two layers of CNOT gates are used to propagate a quarter of the weight-3 stabilizers and gauge operators onto the relevant measure qubit. This information is then measured, and the qubit reset into the appropriate basis. The CNOT layers are then applied again in reverse order to propagate this new Pauli operator back onto the original weight-3 footprint. 

For the unpaired weight-3 stabilizers on the boundary the detectors are formed by pairing consecutive measurements of the stabilizer, while for weight-6 superstabilizers the detectors are formed by consecutive measurements of the two weight-3 gauge operators which make up the superstabilizer. We note that since the measure qubits in this circuit are contributing to measurements of the four stabilizers around them, detectors are not formed between consecutive measurements of the same qubit, but consecutive measurements of the same operator four rounds apart. This is similar to the Surface-13 construction in Ref.~\cite{tomita2014low}, where the bulk measure qubits adjacent to the boundary stabilizers serve double duty, measuring the boundary stabilizers and the bulk stabilizers in alternating rounds. This structure is also seen in Ref.~\cite{leroux2024snakesladdersadaptingsurface}.

To allow a single measurement qubit to measure components of four stabilizers, we need to interleave operations in time. This has an impact on the timelike distance of the circuit. Each stabilizer in the bulk is formed by a pair of gauge operators, meaning it takes two rounds to fully extract the stabilizers for a single basis. After the measurements layers, the codestate is projected into a surface code state, however due to the subsystem nature of the extraction the surface code state has randomized signs for some stabilizers in each round. To extract both bases of information using the same qubit, the stabilizers also alternate position over the course of the circuit. While the boundary types stay consistent throughout, the checkerboard alternates colors in time so that information can flow towards the remaining measurement qubits. This can be seen in the end-cycle states of Fig.~\ref{fig:detslices}. This shifting of information throughout the circuit allows it to efficiently use its measure qubits, but also causes the detecting regions in the circuit to be larger, in both space and time, than those of the usual surface code circuit. This leads to higher detection event fractions and logical error rates.

The primary advantage of these circuits is their component efficiency in comparison to the usual surface code. At large distances, the usual surface code circuit has $2d^2$ qubits and $4d^2$ couplers, as each qubit in the bulk attaches to four couplers, while each coupler attaches to two qubits. For a diamond circuit surface code the qubit count is reduced to $1.5d^2$, and the coupler count ends up being just $2d^2$ since data qubits in this circuit only talk to their two neighbors. In total, this amounts to more than a 40\% reduction in control lines relative to the usual circuit of the given distance, assuming each qubit and coupler utilizes a single line. These counts are only for qubit and coupler control lines, not readout, where the diamond circuits would again have an improvement. Reductions in line count allow us to implement a larger distance using the same number of lines, which is important if we are limited by the number of control lines entering our dilution refrigerator. In addition, it is reasonable to assume that removing two of the couplers on every data qubit would improve noise parameters. For the sake of simplicity, we do not assume any experimental advantage for the numerical results in this paper, so we expect the results in this paper to undersell the advantage of diamond circuits.

\begin{figure}
    \centering
    \includegraphics[height=0.75\linewidth]{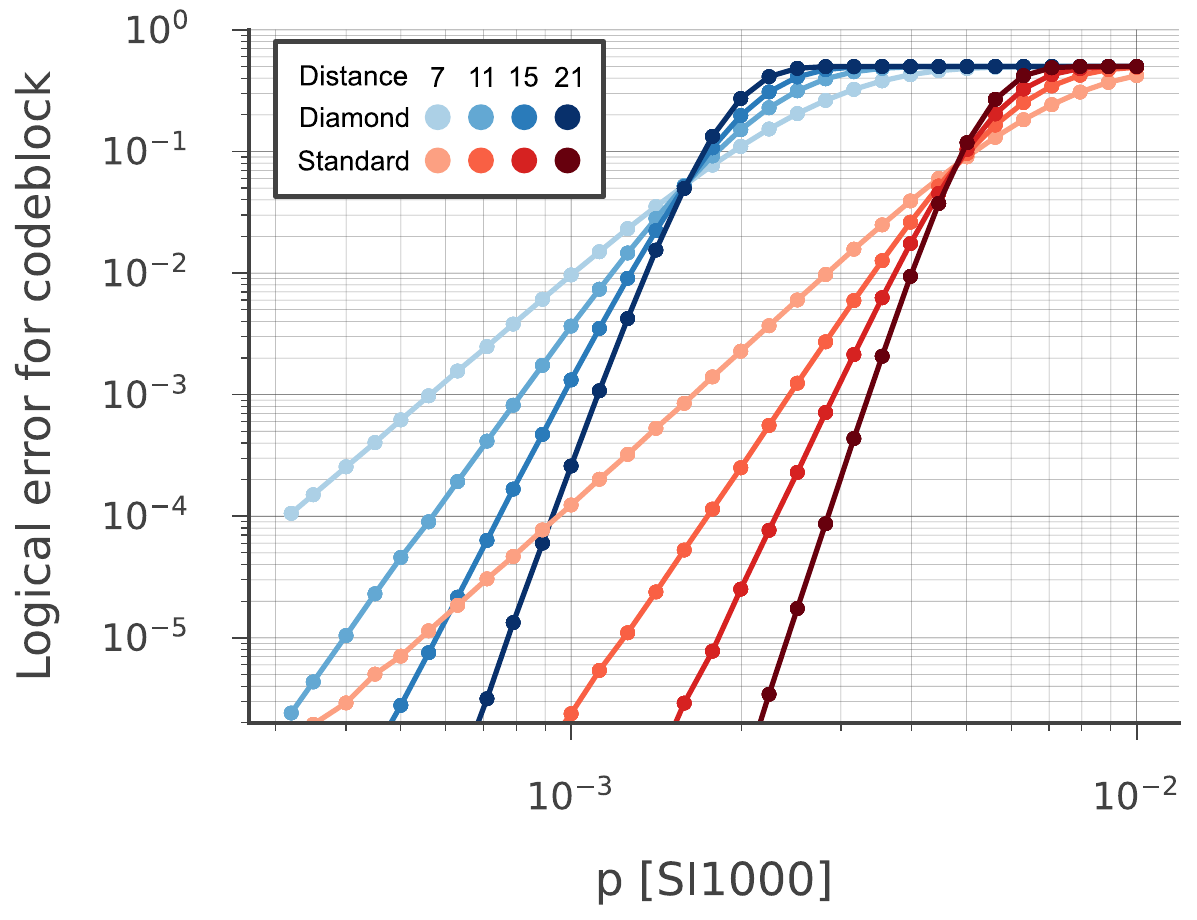}
    \caption{\textbf{Logical error rate comparison. }In this plot we show logical error rates for the standard $d\times d\times d$ blocks of the standard surface code circuit, and for $d \times d \times 4d$ for the diamond circuits, using SI1000 noise~\cite{gidney2022benchmarking}. The points where different distances cross is the threshold for that circuit, with the diamond circuit threshold being roughly $3\times$ lower.}
    \label{fig:ler}
\end{figure}
\begin{figure}
    \centering
    \includegraphics[height=0.755\linewidth]{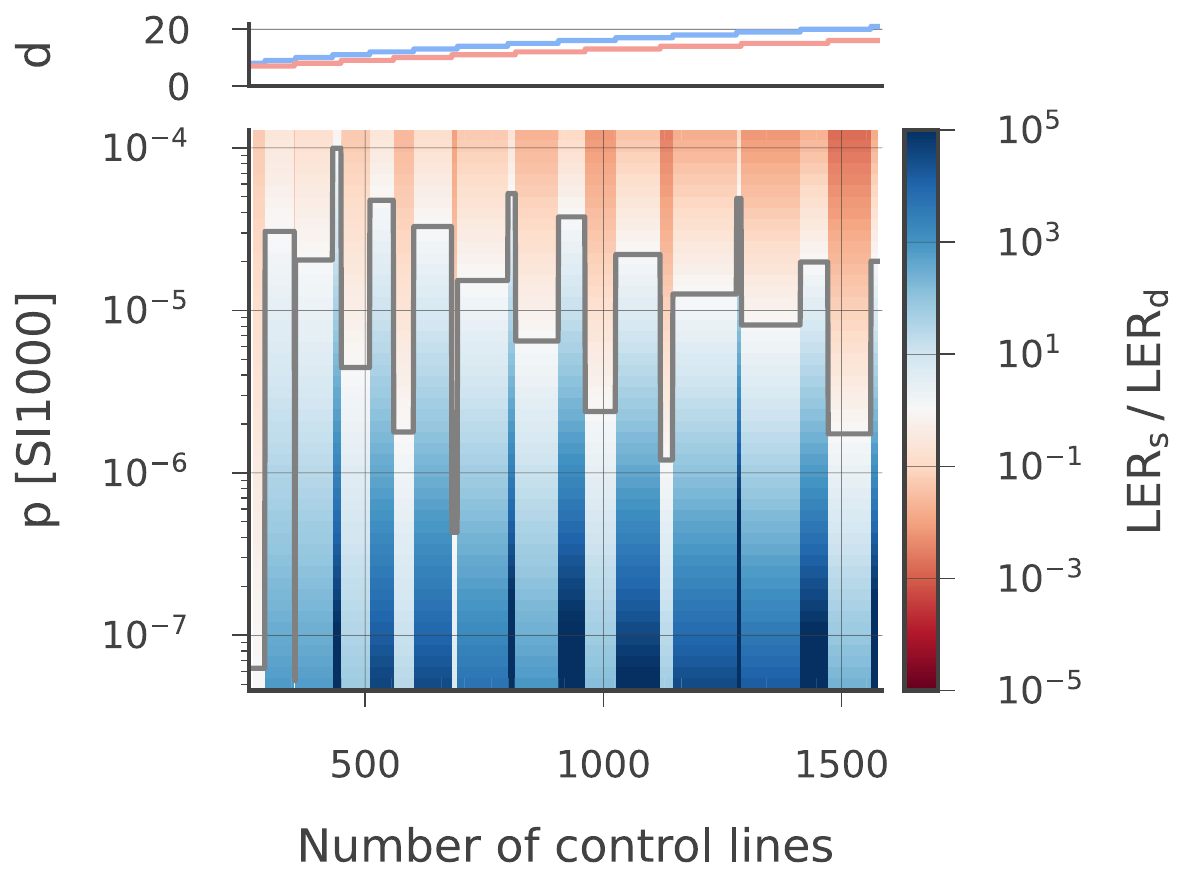}
    \caption{\textbf{Region of when to use each circuit. }The top plot shows the achievable distance for the diamond circuit (blue) and standard surface code circuit (red) for a number of control lines. The bottom plot plot shows in what parameter regimes it is beneficial to use a diamond circuit as opposed to the regular surface code circuit. The grey line shows the error rate below which one should switch to the diamond circuit, for a given number of usable control lines. The discontinuities correspond to changes in code distance for one of the two methods. The background color gives the ratio of logical error rate per codeblock for each circuit.}
    \label{fig:phase diagram}
\end{figure}

In Fig.~\ref{fig:ler} we show the logical error rate performance of the diamond and standard surface code circuits~\cite{dennis2002topological, fowler2012surface, acharya2024quantum}. Since the detecting regions last four times as long as the regular surface code circuit, we compare the logical error rate of a $d\times d\times d$ block of the usual circuit to the logical error rate of a $d\times d\times 4d$ block of the diamond circuits, where a round is defined as a chunk of circuit between two measurement layers. We use SI1000 noise models, as defined in App.~\ref{app:si1000}, simulate circuits using \texttt{stim}~\cite{stim}, and decode using a two-pass correlated sparse blossom matching decoder~\cite{fowler2013optimalcomplexitycorrectioncorrelated, higgott2025sparse}. It is clear that the diamond circuits are not as performant as the regular surface code circuits for a given distance, due to their large detecting regions and more challenging decoding graph. The threshold of the diamond circuits is roughly $3\times$ lower than that of the usual circuit. This means that for a given physical error rate below both thresholds, the diamond circuits have a roughly $3\times$ smaller factor of improvement when increasing distance by two. However when considering line count or qubit count as the scale parameter instead of distance, the diamond circuits are able to win out at sufficiently low error rates. This can be seen in Fig.~\ref{fig:phase diagram}, where we plot the curve of error rates at which it is beneficial to run a diamond circuit instead of the standard circuit.

The structure seen in the gray curve in Fig.~\ref{fig:phase diagram} is due to the discrete jumps in distance that occur at different times between the two circuit families. In the top subfigure, we plot the maximum distance of each family for a given linecount. The upward jumps in the crossover curve correspond to points where the diamond circuit distance increases, while the downwards drops correspond to moments the standard circuit distance increases. Due to the relative ratio of linecounts, these events happen in a repeating pattern. To show this further, we plot the ratio between distances in App.~\ref{app:distance-ratio} and show it asymptotically approaches $\sqrt{6/3.5} \approx 1.31$. The background color on the plot indicates the ratio of codeblock error rates, showing that the transition becomes sharper at larger system scales.

In this manuscript we have presented a family of surface code circuits which measure all the surface code stabilizers while using fewer hardware components. Such an architecture allows for higher-distance circuits relative to the usual surface code for a fixed number of control lines. In addition, experimentally relevant error models like crosstalk during single- and two-qubit gates would be reduced due to additional isolation. If these benefits are significant enough, it is possible that a fault-tolerant quantum computer using diamond circuits could outperform one using a more conventional surface code approach. In contrast with the hex-grid surface codes of Ref.~\cite{mcewen2023relaxing}, which are circuits in the LUCI family with the maximum number of coupler dropouts, these circuits represent an extremal case where the maximum number of qubits have been dropped out while still preserving distance. This result shows that the LUCI framework contains architecturally interesting beyond considering dropout. In addition, it suggests that useful circuits may come from studying the mid-cycle variants of other topological subsystem codes~\cite{Suchara_2011}.

\section{Acknowledgements}
The author would also like to thank Matt McEwen, Craig Gidney, Noah Shutty, and Adam Zalcman, for their help in developing the LUCI framework and Matt McEwen, Alexis Morvan, and Alec Eickbusch, for pointing out the importance of line-efficiency. They would also like to thank Oscar Higgott for his help in improving the manuscript.

\bibliographystyle{apsrev4-1_with_title}
\bibliography{References}
\onecolumngrid
\appendix
\clearpage

\section{SI1000 Noise Model}\label{app:si1000}
The noise model used in this paper is SI1000, a superconducting inspired noise model which assumes a $1000 ns$ cycle time. It is a single parameter noise model defined as follows:

\begin{table}[ht]
    \centering
    \resizebox{\linewidth}{!}{
    \begin{tabular}{|c|l|}
         \hline
         \textbf{Noisy Gate} & \textbf{Definition} \\
         \hline
         $\text{AnyClifford}_2(p)$ & \text{Any two-qubit Clifford gate, followed by a two-qubit depolarizing channel of strength $p$.} \\
         \hline
         $\text{AnyClifford}_1(p)$ & Any one-qubit Clifford gate, followed by a one-qubit depolarizing channel of strength $p$. \\
         \hline
         $\text{R}_{Z}(p)$ & Initialize the qubit as $\ket{0}$, followed by a bitflip channel of strength $p$. \\
         \hline
         $\text{R}_{X}(p)$ & Initialize the qubit as $\ket{+}$, followed by a phaseflip channel of strength $p$. \\
         \hline
         $M_Z(p, q)$ & Measure the qubit in the $Z$-basis, followed by a one-qubit depolarizing channel of strength $p$, \\
         & and flip the value of the classical measurement result with probability $q$.\\
         \hline
         $M_X(p, q)$ & Measure the qubit in the $X$-basis, followed by a one-qubit depolarizing channel of strength $p$, \\
         & and flip the value of the classical measurement result with probability $q$. \\
         \hline
         $M_{PP}(p, q)$ & Measure a Pauli product $PP$ on a pair of qubits, \\
         & followed by a two-qubit depolarizing channel of strength $p$, \\
         & and flip the classically reported measurement value with probability $q$. \\
         \hline
         $\text{Idle}(p)$ & If the qubit is not used in this time step, apply a one-qubit depolarizing channel of strength $p$. \\
         \hline
         $\text{ResonatorIdle}(p)$ & If the qubit is not measured or reset in a time step during which other qubits are \\ &  being measured or reset, apply a one-qubit depolarizing channel of strength $p$. \\
         \hline
    \end{tabular}
    }
    \caption{
        Modified from \cite{gidney2022benchmarking}. Noise channels and the rules used to apply them.
        Noisy rules stack with each other - for example, Idle($p$) and ResonatorIdle($p$) can both apply depolarizing channels in the same time step.
    }
    \label{tab:noise_gates}
\end{table}

\section{Distance Ratios}\label{app:distance-ratio}
\begin{figure}[h]
    \centering
    \includegraphics[width=0.95\linewidth]{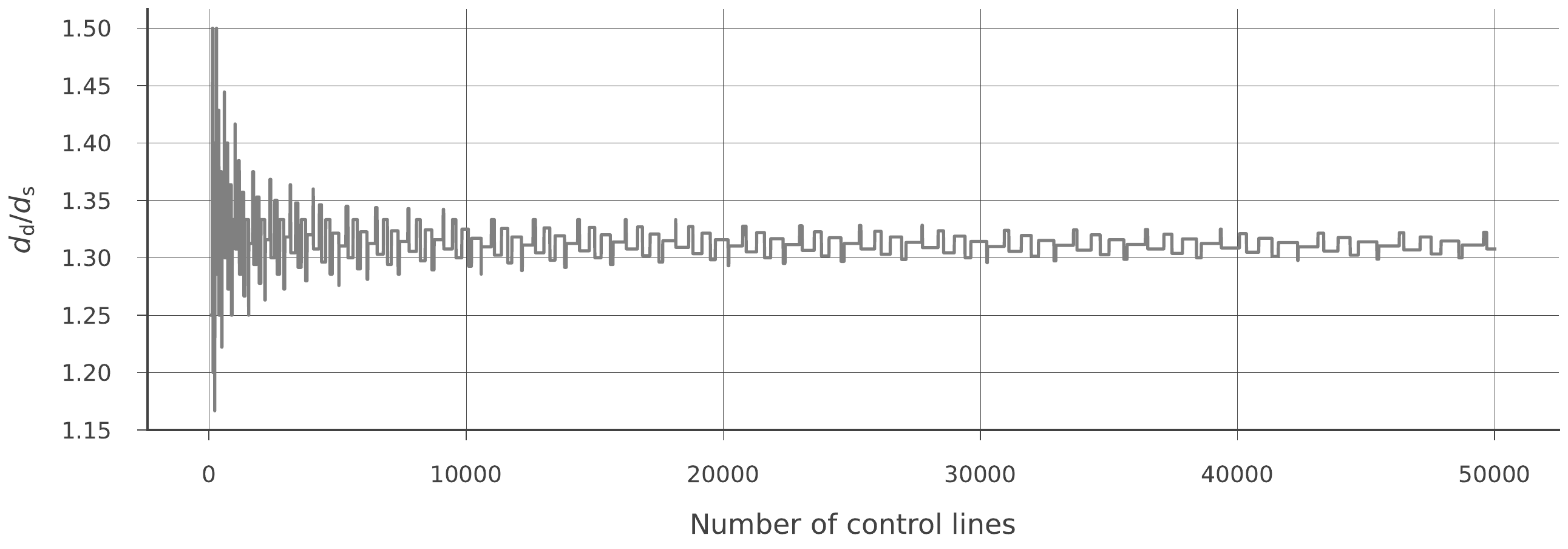}
    \caption{\textbf{Ratio of maximum distances vs. linecount. }The gray curve shows the ratio of distances between the best diamond circuit and best standard circuit for a given linecount. We can see that the ratio follows a regular pattern that matches the curve in Fig.~\ref{fig:phase diagram}.}
    \label{fig:distance-ratio}
\end{figure}
In Fig.~\ref{fig:distance-ratio} we plot the ratio in distances achieved by each circuit for a given system scale. The curve matches up to the curve in Fig.~\ref{fig:phase diagram}, and the repetitive pattern is due to the spacings between $6d^2$ and $3.5d^2$ when $d\in\mathbb{Z}$. As the system gets larger the ratio asymptotes to $\sqrt{6/3.5} \approx 1.31$.

\newpage
\section{Even-distanced diamond circuits}
\begin{figure}[h]
    \centering
    \includegraphics[width=0.95\linewidth]{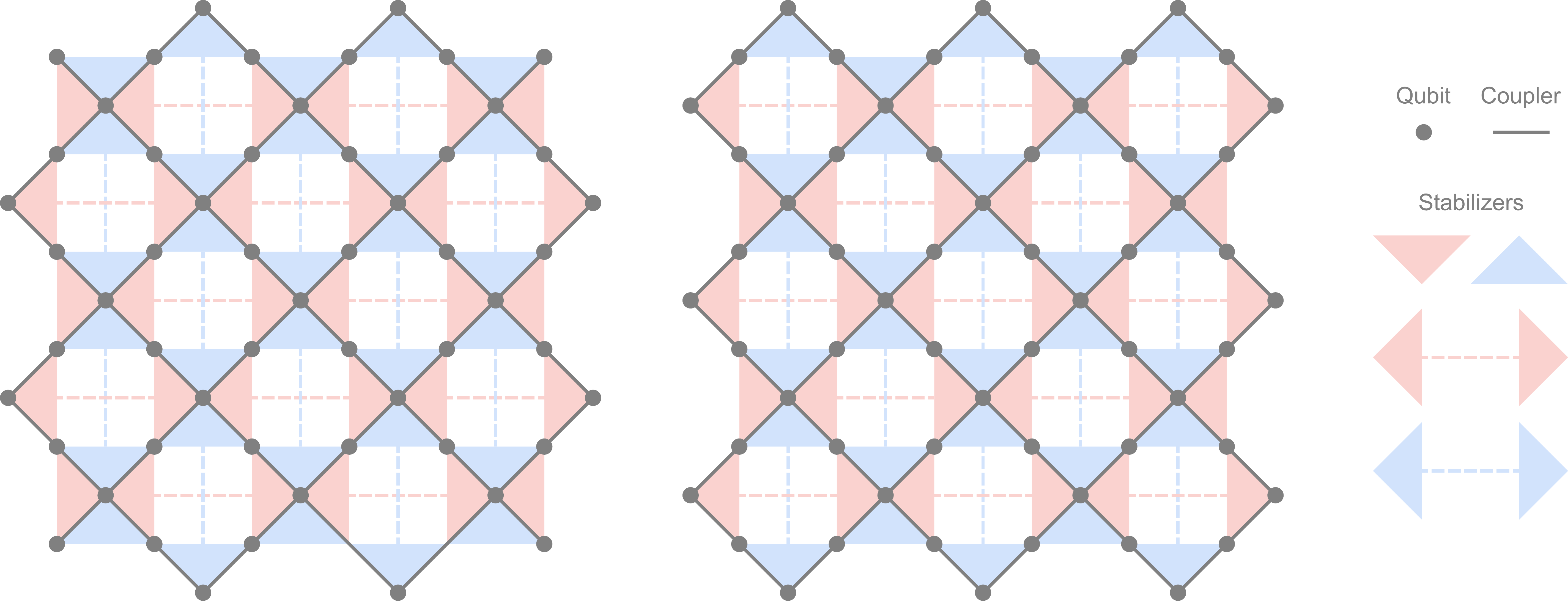}
    \caption{\textbf{Qubit grid and mid-cycle states for distance-6 diamond circuits. }The two variants differ by which subgrid of measure qubits are removed. In contrast with the odd distance diagram in Fig.~\ref{fig:grid}, where two corners touch weight-3 stabilizers and two touch weight-6 stabilizers, the even distance instances have four corners of one type or the other.}
    \label{fig:even-distances}
\end{figure}
For even distance surface codes, there are an odd number of stabilizers along each boundary. Depending on which subgrid of measure qubits are missing, there will be weight-6 or weight-3 stabilizers at each corner. The two circuits have slightly different qubit counts, and performance, but are reasonably similar. In the manuscript, we use even distance circuits from the case on the left, with the weight-3 stabilizers are on the corners. These circuits have fewer qubits, meaning the upwards jumps in the crossover curve of Fig.~\ref{fig:phase diagram} are earlier.

\end{document}